\documentclass{jetpl}
\twocolumn

\lat

\usepackage[version=3]{mhchem} 

\usepackage[dvipsnames]{xcolor}
\usepackage{ulem}
\usepackage{color}

\graphicspath{{pics/}}

\title{Revealing Low-Radiative Modes of Nanoresonators with Internal Raman Scattering}

\rtitle{Revealing Low-Radiative Modes of Nanoresonators\ldots}

\sodtitle{Revealing Low-Radiative Modes of Nanoresonators with Inherent Spontaneous Raman Scattering}

\author{K.\,V.\,Baryshnikova$^{+}$,
K.\,Frizyuk$^+$,G.\,Zograf$^+$, S.\,Makarov$^+$, M.\,A.\, Baranov$^+$, D.\,Zuev$^+$, V.\,A.\,Milichko$^+$, I.\,Mukhin$^{+*}$,M.\,Petrov$^+$, and A.\,B.\,Evlyukhin$^{+V}$\/\thanks{e-mail: k.baryshnikova@metalab.ifmo.ru}}

\rauthor{K.\,V.\,Baryshnikova,
K.\,Frizyuk, G.\,Zograf \dots}

\sodauthor{Baryshnikova,
Frizyuk, Zograf, Makarov, Baranov, Zuev, Milichko, Mukhin, Petrov, Evlyukhin}

\address{$^+$Department of Nanophotonics and Metamaterials, ITMO University, St.~Petersburg, Russia\\~\\
$^*$Laboratory of Renewable Energy Sources, St. Petersburg Academic University, St.~Petersburg, Russia\\~\\
$^V$Institute of Quantum Optics, Leibniz Universität Hannover, 30167 Hannover, Germany}

\dates{18 April 2019}

\abstract{Revealing hidden non-radiative (dark) of resonant nanostructures using optical methods such as dark-field spectroscopy often becomes a sophisticated problem due to a weak coupling of these modes with a far-field radiation, whereas methods of dark-modes spectroscopy, e.g. cathodoluminescence or elastic energy losses, are not always convenient in use. Here, we suggest an approach for experimental determining the mode structure of a nanoresonator basing on utilizing intrinsic incoherent Raman scattering. We theoretically predict the efficiency of this approach and realize it experimentally for silicon  nanoparticle resonators possessing strong Raman line at 520 cm$^{-1}$. With this method we studied  a silicon nanoparticle placed on a gold substrate  and reveal the spectral position of a low-radiative magnetic quadrupole mode which is hardly observable with common dark-field optical spectroscopy.}


\begin{document}

\maketitle

\section{Introduction} 
The resonantly enhanced optical interaction of light with metallic, dielectric,  or semiconductor nanoparticles, which are the basic components of various nanooptical devices, is at the forefront of modern nanophotonics~\cite{Kuznetsov2016, Schuller2010, Staude2017, evlyukhin2012demonstration}. The examples of vital  applications include on-chip integrated photonic systems~\cite{Pernice2015, Pernice2016, Liu2016}, lab-on-chip technologies~\cite{Schwarz2014}, medical biosensing~\cite{Brolo2012} and others. However, understanding the modal content of resonant nanostructures is non-trivial problem due to their substantially subwavelength sizes and typically complicated geometries. This problem has very limited solution within the common optical spectroscopy methods as some of the modes can be low-radiative because of their weak coupling to freely propagating waves, and in this sense are often known as {\it dark-modes}~\cite{Krenn_2012, magdark_5, Yang2015}. Thus, studying the modal spectrum of optical nanostructures with account on dark modes immediately becomes a complicated problem.

\begin{figure}[h]\centering
\includegraphics[width=0.49\textwidth]{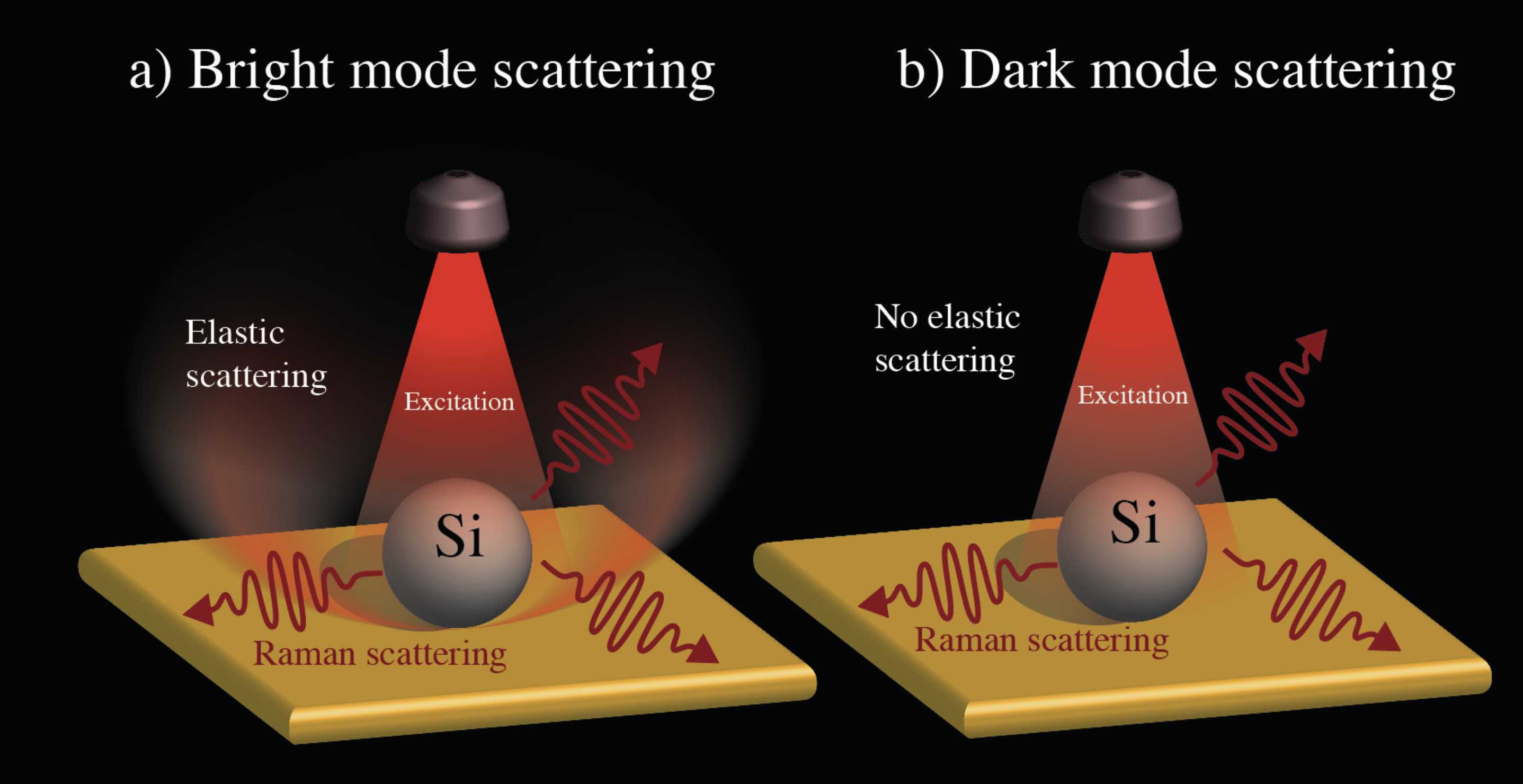}
\caption{The elastic and inelastic scattering by bright and dark resonant modes in silicon nanoparticle on the top of gold substrate. (a) The excitation of the bright nanoparticle mode results in both elastic scattering and inelastic (Raman) scattering. (b) The excitation of the dark mode does not contribute to elastic scattering, but can be detected via Raman scattering signal. }
\label{fgr:scheme}
\end{figure}

The solution of this problem usually relies on sophisticated optical methods such as total internal reflectance microscopy (TIRM)~\cite{Yang2010} and its combination with atomic-force microscopy~\cite{Lahiri2013, Chae2015}. 
The methods based on resonant mode excitation through accelerated electron scattering showed their efficiency in observing the dark-modes of nanostructures. In particular, the elastic electron loss spectroscopy (EELS) is very powerful method for detecting the modal structure of plasmonic resonators by measuring both radiative and non-radiative losses of electrons interacting with the electron excitations in the nanostructures~\cite{Chu2009, Koh2011}. The cathodoluminescence (CL) method similarly to EELS uses accelerated electrons, but excites the dark mode via CL radiation generated inside a nanoresonator. In CL technique, the electronically excited luminescence signal inside the nanostructures is enhanced by the resonant modes, allowing spatial and spectral mapping of dark modes in both plasmonic~\cite{Koenderink_2011, Barnard2011} and non-metal~\cite{Coenen2013, VandeGroep2016, Polman_2016} structures. 

In this Letter, we suggest detecting an intrinsic Raman signal, which corresponds to inelastic photon scattering via optical phonons excitation in a crystalline lattice of a nanostructured solid to determine its electromagnetic modal structure. The Raman signal generated by lattice vibrations is distributed over the volume of the nanoresonator and effectively couples to all resonant modes including dark ones. By exciting the optical phonons and their consequent spontaneous decay, the incoherent Stokes signal is generated inside the nanoresonator under study. The intensity of the Raman signal is governed by the optical density of states inside the nanoresonator at the frequency of the Raman emission, and additionally enhanced at the excitation frequency by resonant excitation of the eigenmodes. We show that the approach allows to reveal magnetic quadrupole mode in a rather complicated system of a silicon nanoparticle arranged on a gold substrate as well as an anapole state, demonstrating the applicability of our approach even when dark-field optical spectroscopy methods are not efficient. Silicon as one of the main materials used in all-dielectric nanophotonics~\cite{Kuznetsov2016, Krasnok2015,  Staude2017} has a well pronounced peak in Raman spectrum at $\Omega \approx 520$~cm$^{-1}$, which is defined by the light coupling with optical phonons at the band central point~\cite{Dmitriev2016a, Yu2010, Cao2006}. 
The  system under study consists of a subwavelength silicon  nanoparticle (SiNP) placed over a gold substrate as show in Figure~\ref{fgr:scheme}.  We note that the effect of a metallic substrate on Mie resonances of SiNP has recently attracted much attention due to the effective optical bi-anisotropy of such system~\cite{Sinev2016, Evlyukhin2015, Miroshnichenko2015, Markovich2014}, surface plasmon polariton launching\cite{Krasnok2018, sinev2018near} and biosensing\cite{krasilin2018conformation}. It should be stressed that approach can be applied to detect any resonant mode excited in any structures that have inherent Raman scattering at visible range.  
\section{Results and discussion}
\begin{figure}[h!]\centering
	\includegraphics[width=0.35\textwidth]{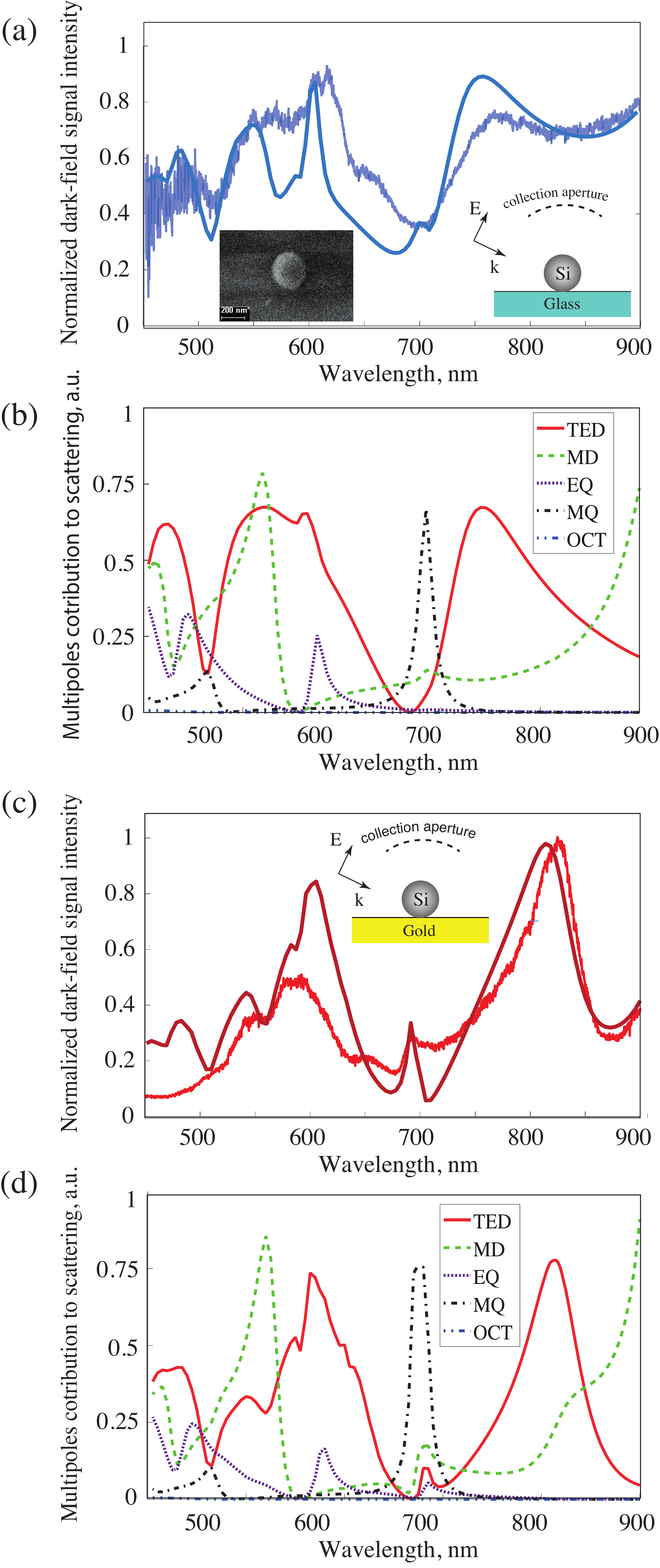}
	\caption{(a)~Experimentally measured and numerically calculated dark field spectra of nanoparticle with radius 125~nm placed on glass substrate. The angle of incidence is 65$^\circ$. The collection aperture and the SEM image of silicon nanoparticle placed on the glass substrate are shown in figure by insets.  (b) Numerically calculated multipoles contributions to scattering of nanoparticle with radius 125~nm placed on the glass substrate. (c) The same as for (a) but for case of gold substrate. (d) The same as for (b) but for case of gold substrate.}
	\label{fgr:elasticsca}
\end{figure}

We started our consideration with the spectral properties of elastic scattering on the spherical nanoparticles.
The nanoparticles of different radii $R=70-140$~nm were fabricated by using laser ablation technique~\cite{Zywietz2014, Dmitriev2016} and then deposited on glass and gold substrates. The distances between neighboring particles was much larger than the excitation wavelength in order to avoid their interaction. Than their geometrical parameters were investigted with scanning electron microscopy (SEM). Thus, SEM picture of a nanoparticle with radius about 125 nm is shown at inset in Figure~\ref{fgr:elasticsca}(a).  Next, the dark-field spectral measurements were performed in the geometry of oblique incidence, when the first objective (Mitutoyo~M~Plan~Apo NIR~x10~0.26~NA) is used to illuminate the sample by polarized light at the angle of $\approx$67$^{\circ}$ to the surface normal with broadband light source (HL-2000 halogen lamp), and the second one (Mitutoyo M Plan Apo~NIR~x50~0.42~NA) is used to collect the scattered light. Inelastic Raman scattering on the SiNPs was studied on the same setup but with the He-Ne laser as a light source with a wavelength of 632.8~nm and intensity $I_0 \approx 1mW/\mu m^2$ which was cut by a notch filter in collected light. The scattered signal subsequently was analyzed by a confocal setup with a spectrometer (HORIBA LabRam HR) and a cooled charge-couled device (CCD) Camera (Andor DU 420A-OE 325) equipped with a 150~g/mm (dark-field experiments) and a 600~g/mm (Raman scattering) diffraction gratings. The position of the nanoparticles was controlled optically via camera (Canon 400D).  The measured (solid curves) typical dark-field spectra along with numerical modeling (dashed curves) are shown in Figure~\ref{fgr:elasticsca}(a) for nanoparticle of radii 125~nm placed on glass substrate and in Figure~\ref{fgr:elasticsca}(b) for nanoparticle placed on gold substrate. Multipole decompositions for both types of substrate are shown also in Figure~\ref{fgr:elasticsca}(c,d).

We note that the optical response of nanoparticle on gold substrate significantly differs from the case of nanoparticle in vacuum or on glass substrate due to a strong coupling with the metal substrate~\cite{Sinev2016}. This is clearly seen after calculating multipoles excited in nanoparticles~\cite{Markovich2014}. One can see that the electric dipole mode gives a predominant contribution to the dark field signal, and its resonances are coincide with the peaks of dark field spectra. At the same time, in this geometry the magnetic mode gives small contribution into the scattering. Magnetic quadrupole (MQ) moment, which was calculated numerically using Comsol Multyphysics software, is of very high-quality in the vicinity of 700~nm, however its presence on dark field spectra is almost neglected.  The minimum in the scattering in the vicinity of MQ resonance can be explained with the multipole decomposition spectra; as one can see, other multipoles are also suppressed in this region, especially electric dipole, which corresponds to the case of so-called anapole state\cite{Miroshnichenko2015, Grinblat2016, Papasimakis2016}. Weak scattering of MQ stems from high quality factor and absorbance in silicon, which is reflected in the spectrum of the electromagnetic energy stored inside the nanoparticle but not in the scattering spectra. Recently, the strong field enhancement at the anapole frequency was utilized for observing a number of effect such as third harmonic generation~\cite{Grinblat2016, Grinblat2016a} or nanolasing~\cite{gongora2017anapole, baryshnikova2019optical}. 
The enhancement of electric energy inside the nanoparticle at the MQ mode with anapole state has a key importance for Raman signal enhancement. Indeed, the Raman scattering can be described semi-classically by introducing the Raman tensor $\hat{R}_{k}$~\cite{Yu2010}. The subscript index $k=1,2,3$ corresponds to three independent polarizations $x,y$ and $z$ of phonon modes corresondingly, which are mutually incoherent. Then polarization vector at the point ${\bf r}$, which generates Raman signal emission will be as follows~\cite{Yu2010}:
\begin{equation}
{\bf p}_{R_{x,y,z}}({\bf r},\omega)=\hat{R}^{i,j}_{k}\ {\bf E}_{0} (r,\omega),
\end{equation}
where ${\bf E}_{0} (r,\omega)$ is exciting field inside the nanoparticle, and the Raman tensor has the following non-zero components:
\begin{equation}
\hat{R}^{2,3}_{1}=\hat{R}^{3,2}_{1}=\hat{R}^{1,3}_{2}=\hat{R}^{3,1}_{2}=\hat{R}^{2,1}_{3}=\hat{R}^{2,1}_{3}=R_0
\end{equation}

where $R_{0}$ is the Raman tensor amplitude.

\begin{figure}[h!]\centering
\includegraphics[width=0.35\textwidth]{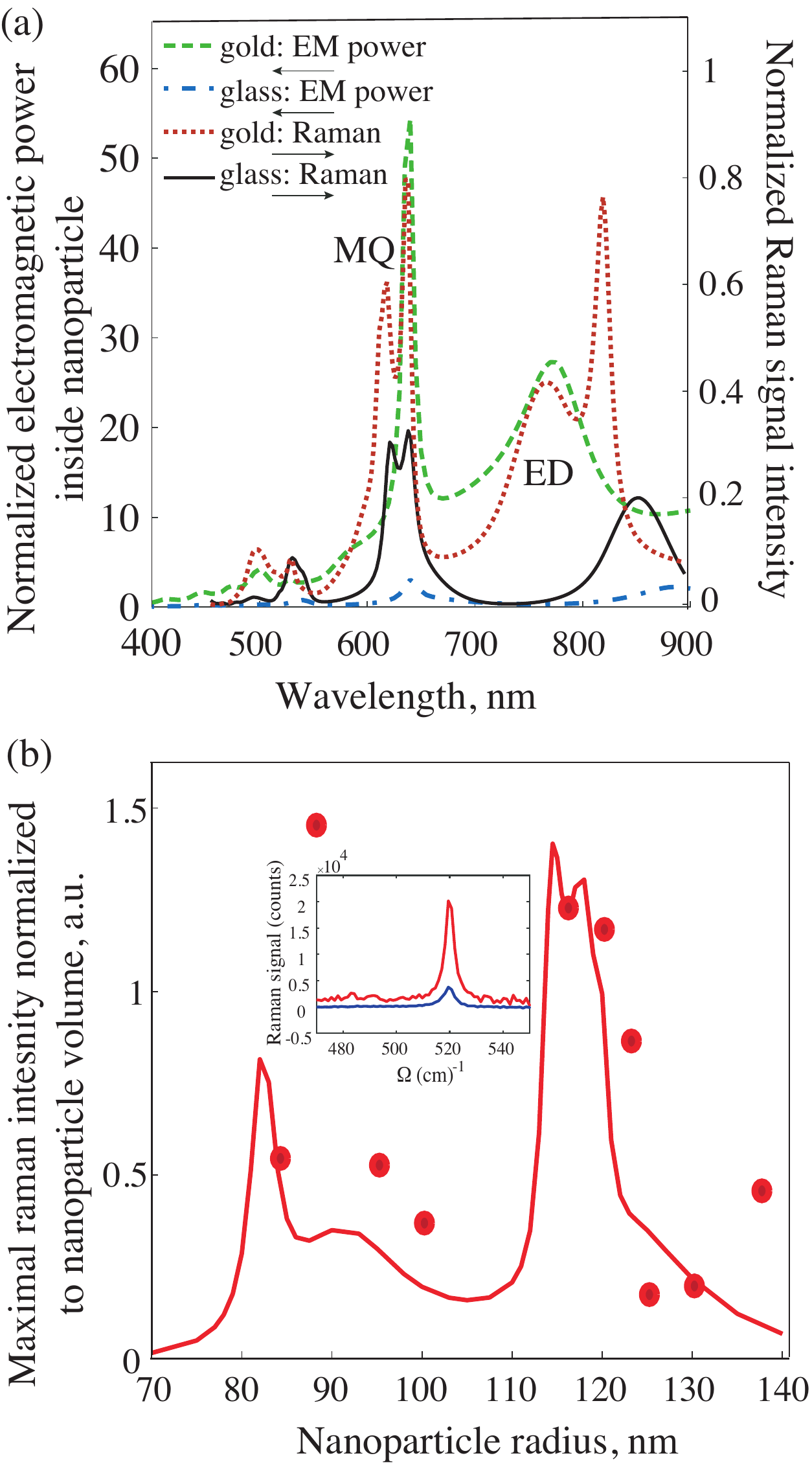}
 \caption{(a)~The modeled spectra  of electromagnetic energy inside the nanoparticle with radius $R=115$~nm placed on gold (dashed red line) or glass (solid blue line) substrate. The modelled dependencies of Raman scattering are shown as function of the excitation wavelength with scattered marks. Inset: typical Raman scattering spectra for a single SiNP placed on glass (blue line) or gold (red line) substrate. (b)~The maximal Raman intensity measured from SiNP of different diameters placed on gold substrate at 633~nm excitation wavelength (red dots). The red solid line corresponds to numerical simulations.}
 \label{fgr:RamanSim}
\end{figure}

The Raman emission is a spontaneous process, and, thus, it is defined not only by the amplitude of Raman polarization, but also by local density of electromagnetic states, which is known as Purcell effect~\cite{Purcell1946,Krasnok2015a}. Then the Raman intensity \cite{frraman} 
\begin{equation}
I_{s}\sim \sum_{s=x,y,z}\int \left|{\bf p}_{R}^{s}(r,\omega)\right|^{2}F_{p}({\bf r},\omega_{S})d\textbf{r}, 
\end{equation}
where $F_{p}$ is the Purcell factor calculated at the frequency of Stocks line $\omega_{S}=\omega-\Omega$, which is lower than the exciting frequency $\omega$ for the frequency of optical phonon $\Omega$, the integral is taken over the whole volume of nanoparticle. The considered SiNP over gold substrate manifest itself as an open resonator, and the Purcell effect in such systems has been intensively discussed  lately~\cite{Zambrana-puyalto2015, Krasnok2015a, Sauvan2013a}. We have used numerical simulations in Comsol Multyphysics$^{\rm TM}$ package to  model the Raman emission spectrum. The results for SiNP of 115~nm radius are plotted in Figure~\ref{fgr:RamanSim} (a) for different excitation wavelengths along with the electromagnetic energy spectrum. Nanoparticle radius was chosen for realization of MQ resonance at 633 nm, which corresponds to the excitation wavelength in the experiment. It should to be stressed that wavelength of MQ resonance stays almost the same for both cases of glass and gold substrate. One can see that the Raman emission intensity generally follows the electromagnetic energy and is enhanced by the resonances of the SiNP. In particular, it is strongly enhanced at the magnetic quadrupole resonance. However, the peak of the Raman emission resonance at the MQ resonance is split as the Purcell factor enhancement is shifted for optical phonon frequency relatively to the peak of the exciting field enhancement at $\omega=\omega_{MQ}$. For this spectral range the optical phonon frequency $\Omega=520 cm^{-1}$  corresponds to $\sim 20$~nm spectral shift for HeNe 633~nm laser excitation, (520cm$^{-1}$), which is comparable to MQ width, that provides the splitting of the Raman intensity peak. This effect is smeared for electric dipole mode which is spectrally much wider than the Raman shift. Herewith total electromagnetic power inside the nanoparticle placed on glass substrate is much smaller than this one for case of gold substrate. The same behaviour is observed for simulated Raman signal. It origins from the high enhancement of fields inside the resonator in the vicinity of metal substrate and the following effect of bianisotropy. 

To verify these assumptions experimentally, we measured the cold Raman emission intensity from SiNP of different sized and arranged on a top of gold substrate. The typical Raman spectrum from silicon nanoparticle placed on gold and glass substrate with sharp maximum around 520~cm$^{-1}$ is shown in Figure~\ref{fgr:RamanSim}(b) by inset which fully prove our assumptions. The excitation wavelength used in the equipment was fixed at 633~nm not allow making spectral sweep of the maximal Raman intensity as shown in Figure~\ref{fgr:RamanSim}(a). However, in order to study the effect of Raman enhancement at different resonances we considered a set of nanoparticles of different radii $R=70-140$~nm. Due to the properties of Mie resonances their resonant wavelengths shift linearly to the long wavelength region with increasing nanoparticle size and the multipole resonances coincide with the excitation wavelength one after the another. Finally, for different size of SiNP we obtained dependence shown in Figure~\ref{fgr:RamanSim}(b), which is obtained after normalizing the Raman intensity over nanoparticle's volume. The measured experimental data (red circles) show a good correspondence to the results of numerical modeling (red solid line). The important result here is that the specific contribution of MQ mode in Raman intensity is much higher than for dipole resonances, contrary to the case of elastic scattering, when the MQR is almost undetectable in total scattering spectra.



\section{Conclusions}
In conclusion, we have suggested an efficient approach for experimental determining the spectral content of nanoresonator basing on utilizing an incoherent Raman scattering, inherent to the resonator material. The approach was confirmed  theoretically and  experimentally for silicon nanoparticle resonators possessing a strong intrinsic  Raman response. We have shown that the suggested method allows us to reveal the magnetic quadrupole resonance combining with the anapole state in the system of a silicon nanoparticle arranged on a gold substrate, demonstrating the applicability of our approach for broad range of non-plasmonic resonant nanostructures. 

Numerical simulations and ivestigation of anapole states was financially supported by Russian Science Foundation (project 17-72-10230). The experimental part was supported by Russian Science Foundation (project 18-79-00338).

\end{document}